\documentclass[twocolumn,showpacs,footinbib]{revtex4}

\usepackage{bm,enumerate,dcolumn,tikz,graphicx,color,amsmath,amssymb}
\usepackage{epstopdf}
\usepackage{empheq}
\usepackage{capt-of}
\usepackage{pgfplots}

\usepgfplotslibrary{colormaps,external}
\usetikzlibrary{decorations.markings}
\usetikzlibrary{arrows}

\newcommand{\comment}[1]{}
\newcommand{\bra}{\langle}
\newcommand{\ket}{\rangle}

\definecolor{lightblue}{rgb}{.8, .8, 1}
\definecolor{myyellow}{RGB}{254,241,24}
\definecolor{myorange}{RGB}{234,125,1}
 
\usepackage{array}
\newcolumntype{C}[1]{>{\centering\let\newline\\\arraybackslash\hspace{0pt}}m{#1}}
\newcolumntype{L}[1]{>{\raggedright\let\newline\\\arraybackslash\hspace{0pt}}m{#1}}
\newcolumntype{R}[1]{>{\raggedleft\let\newline\\\arraybackslash\hspace{0pt}}m{#1}}

\begin{document}

\title{
Intensity landscape and the possibility of magic trapping of alkali Rydberg atoms 
in infrared optical lattices
} 
\author{Turker Topcu and Andrei Derevianko}
\affiliation{Department of Physics, University of Nevada, Reno, NV 89557, USA}
\date{\today}

\begin{abstract}
Motivated by compelling advances in manipulating cold Rydberg (Ry) atoms in optical traps, 
we consider the effect of large extent of Ry electron wave function on trapping potentials. 
We find that when the Ry orbit lies outside inflection points in laser intensity landscape, 
the atom can stably reside in laser intensity maxima. Effectively, the free-electron AC 
polarizability of Ry electron is modulated by intensity landscape and can accept both 
positive and negative values. We apply these insights to determining magic wavelengths 
for Ry-ground-state transitions for alkali atoms trapped in infrared optical lattices. We find 
magic wavelengths to be around 10 $\mu \mathrm{m}$, with exact values that depend on 
Ry state quantum numbers. 
\end{abstract}

\pacs{37.10.Jk, 32.10.Dk, 32.80.Qk, 32.80.Ee}

\maketitle

Magic trapping~\cite{YeKimKat08} of cold atoms and molecules is a powerful technique that  has  
recently enabled ultrastable optical lattice clocks~\cite{TakHonHig05,LudZelCam08etal,DerKat11}, 
long-lived quantum memory~\cite{RadDudZha10}, and precision manipulation of ultracold 
molecules~\cite{NeyYanMos12}. 
When neutral atoms are trapped, their  internal energy levels are necessarily perturbed 
by spatially inhomogeneous trapping fields. For a cold atomic cloud, typical mK 
temperatures translates into the 10 MHz  trap depths.  In other words, as the atom travels 
about the trap, its energies are modulated at the 10 MHz level with  associated coherence 
times of just 100 ns. If it were not for magic trapping techniques, such decoherences would 
be prohibitive for the enumerated cold-atom applications. The key idea of magic trapping 
is the realization that one is interested in differential properties of two 
levels, such as the clock frequency or a differential 
phase accumulated by two qubit states. Then if the trapping field affects 
both levels in the very same way, the differential perturbations vanish. 
Such engineered traps are commonly referred to as ``magic''. These ideas  
enabled precision clock spectroscopy at the sub-100 mHz level~\cite{LudZelCam08etal} and 
second-long coherence times~\cite{RadDudZha10}, orders of magnitude better than the quoted 
``non-magic'' values. 

Application of magic trapping techniques to Rydberg (Ry) states of alkali atoms has 
turned out to be challenging. Generic quantum-information protocols involve qubits encoded in 
hyperfine manifolds of the ground state (GS) and conditional multiqubit dynamics mediated by 
interactions of Ry states~\cite{JakCirZol00,UrbJohHen09,GaeMirWil09,SafWalMol10}. 
Therefore,the trapping field must be magic both for the GS hyperfine manifolds and also for the 
GS-Ry transition~\cite{MorDer12}. 
The first part by itself is a non-trivial problem and has been a subject of several 
studies~\cite{Der10DoublyMagic,Der10Bmagic,FlaDzuDer08,ChiNelOlm10,RadDudZha10}. The GS-Ry transition 
presents another challenge~\cite{SafWilCla03,SafWal05,ZhaRobSaf11,MorDer12}. 

To appreciate the problem, let us first review commonly invoked arguments. In optical 
fields, the trapping potential is proportional to the AC 
polarizability $\alpha(\omega)$, leading to trapping potential 
$U(\mathbf{R})= - \alpha(\omega) F^2(\mathbf{R})/4$,  where $F$ is the local value of the electric 
field~\footnote{Unless specified otherwise, atomic units, $|e|=\hbar=m_e\equiv 1$ are used 
throughout the paper}. The GS polarizability $\alpha_{g}(\omega)>0$ 
when red-detuned from atomic resonances, and the atoms are attracted to intensity maxima. 
On the other hand, loosely-bound Ry electron is nearly ``free''; therefore its polarizability 
$\alpha_{r} (\omega) \approx -1/\omega^2$  is negative and the atoms are pushed 
towards intensity minima. Then as the GS population is driven to a Ry state during 
gate operations, an atom experiences time-varying trapping potential. This causes undesirable 
motional heating. The resulting decoherence is so severe that  experimentalists simply turn 
off trapping fields during GS-Ry excitations~\cite{SafWal05,ZhaRobSaf11,SafWalMol10}. This process 
is also detrimental because this again leads to 
heating. Such ``pulsed" operation of the trap is the leading heating mechanism in gate experiments, 
with heating rate of $\sim$1\% per gate cycle~\cite{SafWal05}. In addition, there is 
a problem of scalability: it is impossible to turn off the trapping field at individual 
trapping sites, therefore the entire ensemble becomes heated during pulsed trap 
operations~\cite{ZhaRobSaf11}. 

One way to evade the AC Stark shifts is to use blue detuned bottle beam traps~\cite{IvaIsaSaf13}. 
Here, the atoms are trapped in intensity minima. Experimentally, such a multi-trap setup 
is arguably more challenging than using optical lattices as discussed below. 

Because of the GS-Ry polarizability sign difference, it is usually accepted that 
magic trapping in red-detuned fields is unattainable 
(see however Ref.~\cite{YouKnuAnd10} for evidence to the contrary). 
Here we present clear arguments that Ry atoms can in fact be attracted to intensity maxima, 
and demonstrate trapping of GS and Ry state in red detuned magic infrared lattices. 

\begin{figure}[h!tb]
	\begin{center}
		\resizebox{79mm}{!}{\includegraphics[angle=0]{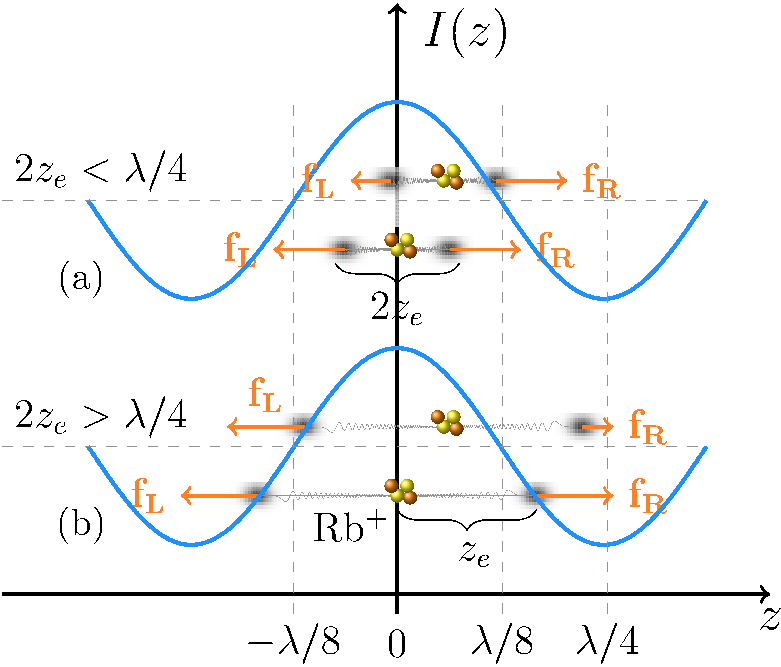}}
	\end{center}
	\caption{(Color online) Influence of Ry orbit size on stability of atomic motion at the 
	1D lattice anti-node for a toy model of Ry atom.
	The  vertical axis is the laser intensity $I(z)$.  Electron cloud is localized in two 
	``lumps''. The  optical dipole force exerted on each lump is directed towards nearby 
	intensity minimum and is proportional to the local intensity slope, $dI(z)/dz$. 
	When the Ry orbit size $2z_e<\lambda/4$ (panel (a)), the $\mathbf{f}_R$ and 
	$\mathbf{f}_L$ acting on the localized ``lumps" of electron density pull the atom away 
	from the intensity maximum (unstable equilibrium). When $2z_e>\lambda/4$ (panel (b)), 
	they act as restoring forces, with Ry atom stably resting at  the intensity maximum. 
	}
	\label{Fig:Toy}
\end{figure}

We start by presenting a qualitative argument (see Fig.~\ref{Fig:Toy}) 
that makes the underlying physics transparent. First of all one realizes that  Ry wave function is 
spread over large distances that can be comparable to spatial  scale of the laser intensity variations.
For example, for a lattice formed by CW lasers of wavelength $\lambda$, 
the laser intensity is spatially modulated with a lattice constant 
$\lambda/2$: $F^2(z)=F^2_0 \sin^2( k z )$, $k=2 \pi/\lambda$.  Ry orbit is larger than the lattice 
constant if its principal quantum number $n > \sqrt{\lambda/(3 a_0)}/2$, e.g., $n \gtrsim100$ for 
$\lambda= 10 \mu\mathrm{m}$. As a result, the ponderomotive potential experienced by the nearly free electron 
must be averaged over local field intensities~\cite{DutGueFel00,YouAndGeo10,ZhaRobSaf11},
\begin{equation}
  U(\mathbf{R}) = \frac{1}{4 \omega^2} \int d^3r_e |\Psi(\mathbf{r}_e)|^2 F^2(\mathbf{R}+\mathbf{r}_e) \, .
  \label{Eq:Uponder}
\end{equation}
Here $\mathbf{R}$ is the position of  atomic core in the laboratory frame, 
$\mathbf{r}_e$ is the Ry electron coordinate relative to the core, and $| \Psi(\mathbf{r}_e)|^2$ is 
the Ry electron probability density. 

An illustration of the interplay between the Ry orbit size and the lattice constant 
can be made for a 1D optical lattice and a toy model of a Ry atom, 
(see Fig.~\ref{Fig:Toy}). Here  $| \Psi(z_e)|^2$ is localized in two ``lumps'' positioned symmetrically 
(at relative distances $\pm z_e$) about the core. The optical-dipole force acting on each lump is transferred to 
the core by the Coulomb interaction, so that the net force on the atom centered at $Z$ reads
\begin{equation}
 f_z( Z) = -\frac{1}{8 \omega^2} \left ( \left. \frac{dF^2}{dz} \right|_{z=Z+z_e}  
			+  \left.  \frac{dF^2}{dz}  \right|_{z=Z-z_e}  \right) \,.
 \label{Eq:forceToy}
\end{equation}
Because of the symmetry,  this force vanishes at the nodes and antinodes of the lattice. 
In Fig.~\ref{Fig:Toy}, we investigate the stability of the atomic motion at intensity maxima. 
It is clear that for small Ry orbits the Ry atom is expelled to a nearby antinode, 
as commonly expected. When $2 z_e = \lambda/4$ the two ``lumps'' sit initially ($Z=0$) at  inflection points 
in the intensity profile and  the atom experiences no net force regardless of displacement. However, 
as the size of the Ry orbit is increased  ($2 z_e > \lambda/4$)  the atom stably resides in intensity 
maxima. As the atom is displaced in some direction the wave 
function lump on the side opposite to the displacement  experiences larger dipole-force tug 
towards nearby node, resulting in  a restoring force.  

From this  model, we can generalize to an arbitrary  3D laser intensity 
landscape: {\it a Ry atom is drawn to intensity maxima if in an equilibrium position (where 
$\mathbf{\nabla}I(\mathbf{R}) = 0$) the bulk of the Ry wave function straddles outside of the nearby 
surface of inflection points in intensity landscape, parametrically given by $\Delta I(\mathbf{r}) = 0$.}
For example, for a 2D Gaussian intensity cone, $I(x,y) = I_0 \exp( - (\rho/\rho_0)^2 )$, the radius of 
Ry orbit must be greater than $\rho_0 ( \sqrt{3}-1)/2 a_0 $. Then a Ry atom can be attracted to the intensity maximum and its motion 
can be guided by the beam. 

Now we explicitly compute the trapping potential. Integrating over ``lumps'' in Eq.(\ref{Eq:Uponder}), 
\begin{equation}
U_t(Z) = \frac{F_0^2}{4\omega^2} \left( 
\sin^2 (k z_e) +
\cos(2 k z_e) \sin^2( k Z) 
\right) \, .
 \label{Eq:ToyPot}
\end{equation}
There are two distinct contributions, $U_t(Z) = U_t^0 + U_t^Z   \sin^2( k Z)$: the $U_t^0$ term is a 
uniform shift across the lattice and the second contribution is 
proportional to the standing wave intensity. The uniform offset does not affect  atomic motion as 
it does not contribute to the force (\ref{Eq:forceToy}). The lattice depth prefactor, 
$U_t^Z=F_0^2/4\omega^2 \cos(2 k z_e)$ shows that the atoms are attracted to lattice nodes if 
$\cos(2 k z_e)>0$, {\it i.e.} when $p -1/4 < 2 z_e/\lambda < p+1/4$, $p=0,1,2,\ldots$. Otherwise, 
atoms reside at antinodes. At critical values $2 z_e/\lambda=(p+1/2)/2$, the trapping potential 
vanishes altogether and the atom travels through the lattice uninhibited. These observations are consistent 
with our dipole force analysis in Fig~\ref{Fig:Toy}.  As we increase the principal quantum number and 
the orbit grows larger, the atoms initially stably reside at nodes. Then as $2 z_e$ reaches  $\lambda/4$, 
the atoms move freely, and then are pushed towards the anti-nodes. This pattern repeats itself with further 
increase in the Ry orbit size. 

Having qualitatively understood the nature of various trapping regimes, we now proceed with a rigorous 
evaluation of trapping potentials for realistic Ry atoms~\cite{OvsDerGib11}.
One starts with the  Hamiltonian $(\mathbf{p}_e -\mathbf{A}/c)^2 /2$,  where $\mathbf{p}_e$ is the electron 
momentum and $\mathbf{A}$ is the vector potential. Upon expanding the square, we encounter the kinetic 
energy term, $\mathbf{p}_e \cdot \mathbf{A}$ cross terms and an $A^2$ contribution. 
It is the $A^2$ term in the Coulomb gauge that leads to the  ponderomotive potential, and in the lowest 
order perturbation theory we recover Eq.~(\ref{Eq:Uponder}). 
Further discussion of the validity of this approximation can be found in~\cite{TopDer13}. 

Explicitly evaluating the integral (\ref{Eq:Uponder}), we find that the Ry atom potential in a 
1D  lattice  is identical to that of our toy problem in 
Eq.~\eqref{Eq:ToyPot}, but with potential shift and depth redefined in terms of expectation values
\begin{eqnarray}\label{eq:pot_zdep}
U^Z_r &=& \frac{F_0^2}{4\omega^2} \bra n l m|\cos (2k z_e)|n l m \ket 
	\equiv- \alpha_r^{\rm lsc}(\omega) \frac{F_0^2}{4} \, , \\
 U^0_r &=&  - \frac{ \alpha_r^{\rm lsc}(\omega) - \alpha_e(\omega)}{2} \,  \frac{F_0^2}{4} \, ,
\label{eq:pot_offset}
\end{eqnarray}
for a Ry state $|n l m\ket$. Here we introduced the effective intensity landscape-averaged polarizability 
$\alpha_r^{\rm lsc}(\omega)=-\bra \cos (2k z_e) \ket/\omega^2$, which unlike the free electron 
polarizability, $\alpha_e =-1/\omega^2$, can accept both positive and negative values. One can 
view $\alpha_r^{\rm lsc}$ as landscape-modulated free-electron polarizability as 
$\alpha_r^{\rm lsc} = \bra \cos(2k z_e) \ket \alpha_e(\omega)$ and 
$|\alpha_r^{\rm lsc} (\omega)| \le |\alpha_e(\omega)|$. 

The optical 
potential is $U_r(Z)=U^0_r + U^Z_r  \sin^2( k Z)$. As in our toy model, the potential consists 
of a term that depends on the position of the atom in the lattice  and a uniform offset $U^0_r$.
Note that without properly accounting for the finite size of Ry cloud, one would 
conventionally write $U_r^{\rm conv} (Z) = F_0^2/ (4 \omega^2) \sin^2(kZ)$. The two potentials, conventional 
and ours, are equal only in the limit $\bra r_e \ket \ll \lambda$ as $U^0_r\rightarrow 0$, and  
$U^Z_r \rightarrow  F_0^2/(4 \omega^2)$ in this limit. Our  potential can support stable equilibrium 
in lattice anti-nodes, while the conventional potential does not. 

The expectation value of $\cos(2kz_e)$ can be evaluated  by expanding $\cos(2kz_e)$ over 
spherical Bessel functions. For example, for $l=0$ states,
\begin{equation}\label{cos_expansion}
\bra ns |\cos(2kz_e)| ns \ket = \int_0^\infty dr_e 
	 P_{ns}^2(r_e)  j_{0}(2kr_e) \, .
\end{equation}
Here  $P_{ns}(r_e)$ is the radial  wave function of Ry electron; 
we computed $P_{ns}(r_e)$ using well-known model potentials~\cite{Gal05book}. 

\begin{figure}[h!tb]
	\resizebox{85mm}{!}{\includegraphics{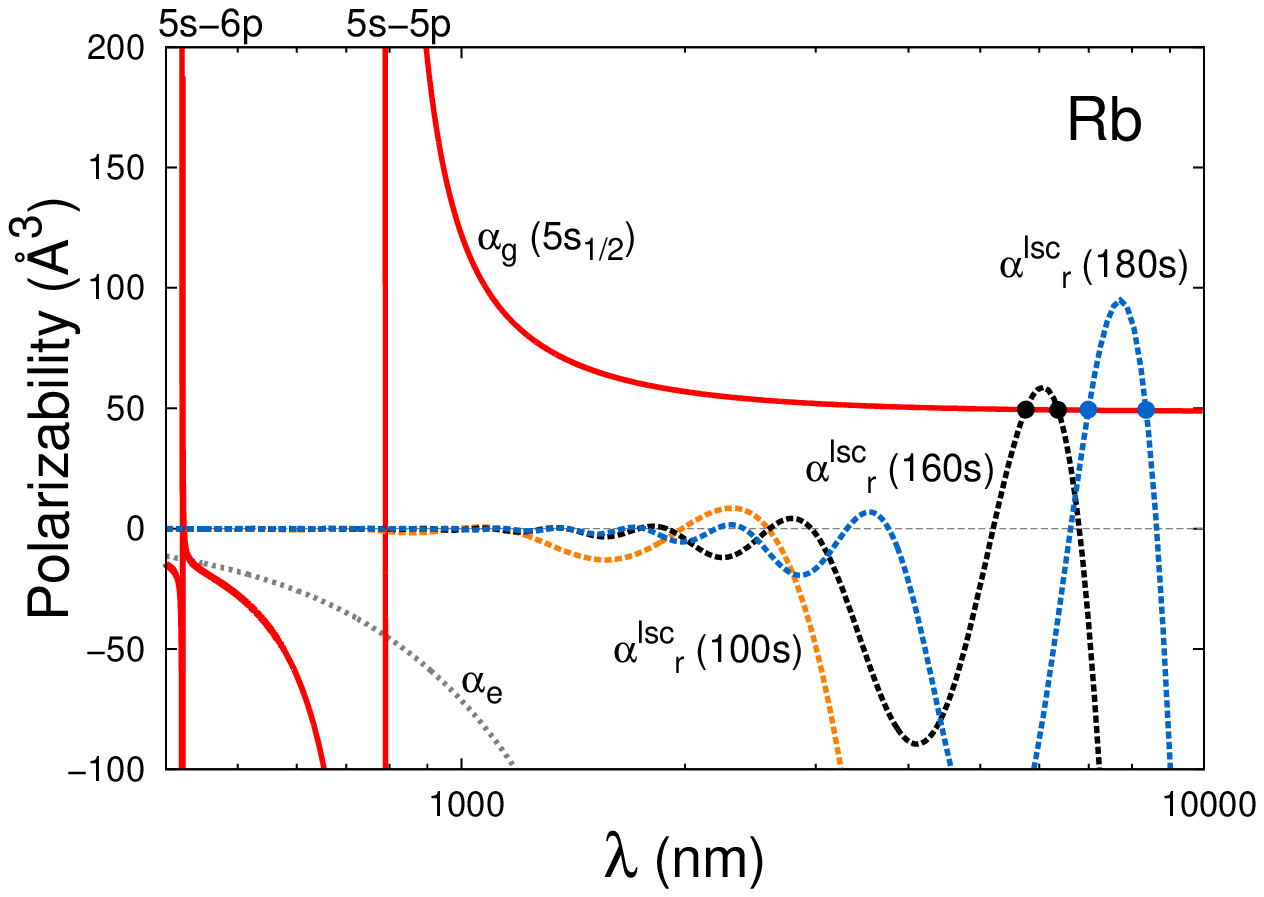}} 
	\caption{The ``landscape-modulated" polarizability $\alpha^{\rm lsc}_{r}$ for $ns$-states 
	of Rb with $n=100$ (dashed orange), 160 (dashed black) and 180 (dashed blue). 
	The scalar ground state polarizability $\alpha_{g}$ (solid red) and the free 
	electron polarizability $\alpha_{e}$ (dashed gray) are also shown. 
	The magic wavelengths can be found for all $n\ge 154$ in the infrared wavelength 
	range spanning the CO$_2$ and the frequency-doubled CO$_2$ laser bands. 
	}
	\label{fig:Rb_magic}
\end{figure}

Our computed landscape-averaged polarizabilities for several Ry states are shown in 
Fig.~\ref{fig:Rb_magic}. 
For all the Ry states, $\alpha^{\rm lsc}_{r}$ is essentially zero at wavelengths below 
$\sim$1000 nm and starts oscillating with increasing amplitude before dropping off like 
$\alpha_e(\omega)$. 

By recasting $\alpha^{\rm lsc}_{r}(\omega) = -( 1 - 2\bra \sin^2(kz_e)\ket )/\omega^2$, one 
could conclude~\cite{OvsDerGib11} that, $\bra \sin^2(kz_e)\ket \approx 1/2$ for 
$\lambda \ll \bra r_e \ket$, so that $\alpha^{\rm lsc}_{r}\approx 0$. 
As $\lambda$ is increased, $kz_e$ gets smaller and $\bra \sin^2(kz_e)\ket\ll 1$ resulting in 
$\alpha^{\rm lsc}_{r} \rightarrow \alpha_e$.  This explains the short- and long-wavelengths 
behavior of $\alpha^{\rm lsc}_{r}(\omega)$ in Fig.~\ref{fig:Rb_magic}. 
As to the $n$-dependence of $\alpha^{\rm lsc}_r$, one could show on general grounds that 
$\bra \cos(2kz_e) \ket=f((n^*)^2 a_0/\lambda)$, where $f$ is some universal function of 
the effective quantum number $n^*$. Thus our discussion is applicable to all Ry atoms. 

A possible detrimental effect on the gate fidelity can arise if the wavefunctions of the Ry 
atoms trapped at adjacent lattice sites start to overlap. In order to avoid such overlaps, 
one can fill in the optical lattice by leaving empty lattice site(s) between trapped atoms. 
Such experimental capabilities have been demonstrated~\cite{Bloch2012}. 
%
Furthermore, the Rydberg blockade mechanism, central for the gate operations relies on the {\em repulsive} long-range interaction~\cite{LukFleCot01}. Penning ionization requires close approach of two atoms, so the rate is suppressed. As experimentally shown in Ref.~\cite{AmtReeGie07},  since the collisional ionization requires attractive potentials, the Ry states first need to be redistributed (mainly by black body radiation~\cite{Gal05book}) to populate Ry states that would correlate to attractive molecular potentials at long range. Therefore  decoherence rates due to ionization is limited from above by the BBR-induced decoherences. BBR-induced decoherence were studied in~\cite{SafWal05}: at room temperature Ry-state lifetime is greater than 0.1 ms for $n>65$. Thereby, collisional ionization being the secondary step, is not an issue for Ry gates.


Now since $\alpha^{\rm lsc}_{r}(\omega)$ can become positive, we show that the GS and Ry 
 potentials can be matched at red-detuned ``magic" wavelengths. For the 
GS atoms, the trapping potential reads
\begin{equation}
 U_{g}(Z) = - \frac{F_0^2}{4} \alpha_{g}(\omega)  \sin^2( k Z) \, ,
\end{equation}
with the dynamic polarizability ($\mathbf{D}$ is the dipole operator and $E_i$ are atomic energy levels)
\begin{equation}\label{dynpol}
\begin{split}
\alpha_{g}(\omega)  = \sum_{i}\;\frac{(E_g-E_i)| 
	\bra\psi_g | \mathbf{D}|\psi_i\ket|^2}{(E_g-E_i)^2-\omega^2}\; . 
\end{split} 
\end{equation} 
We  evaluated  $\alpha_{g}(\omega)$ with a high-accuracy
procedure~\cite{DerJohSaf99}.

The two spatial parts of Ry and GS  potentials  match when $U^Z_r = U^Z_g$, which 
is  attained at ``magic'' values of laser frequencies $\omega^*$ when 
$\alpha_{g}(\omega^*)=\alpha_{r}^{{\rm lsc}}(\omega^*)$.
In Fig.~\ref{fig:Rb_magic} we  plot both polarizabilities 
to search for such magic wavelengths $\lambda^*$. We find 
that for Rb $ns$ states the two curves cross for all $n \ge 154$ with $\lambda^*\simeq 5600$ nm 
for $n=154$. Above this critical value of $n$, there are at least two values of $\lambda^*$ ({\it e.g.}, 
$\alpha^{\rm lsc}_r$ for the 160$s$ and 180$s$ states cross twice with the GS polarizability). 
The number of  $\lambda^*$'s increases further with increasing $n$. Table~\ref{table:magic_wlen} 
compiles $\lambda^*$ for Rb and  Na atoms. In addition to the $l=0$ states, 
Table~\ref{table:magic_wlen} lists  $\lambda^*$ for the $l=1$ and $l=2$ states (all $m=0$). 

\begin{center}
\label{table:magic_wlen}
\begin{figure}[h!tb]
\captionof{table}{Magic wavelengths (nm) for Rydberg states of Na and Rb atoms, $l=0,1,2$ and $m=0$.  }
\begin{tabular}{ C{0.5cm} C{1.15cm} C{1.15cm} C{1.15cm} C{1.15cm} C{1.15cm} C{1.15cm}}
\hline\hline
 &\multicolumn{3}{c}{$n=$100}&\multicolumn{3}{c}{$n=$120}\\
\cline{2-7}
     & $s$ & $p$ & $d$    & $s$ & $p$ & $d$ \\
\hline
Na   & 
	& 1961 \newline 4655 
	& 2798 \newline 3762 
	& 
	& 2734 \newline 6766 
	& 3894 \newline 5532  \\ 
\hline
Rb   & 
	& 2022 \newline 4421 
	& 3016 \newline 3401 
	& 
	& 2750 \newline 6525 
	& 3942 \newline 5301  \\ 
\hline
\end{tabular}

\begin{tabular}{C{0.5cm} C{1.15cm} C{1.15cm} C{1.15cm} C{1.15cm} C{1.15cm} C{1.15cm}}
 &\multicolumn{3}{c}{$n=$160}&\multicolumn{3}{c}{$n=$180}\\
\cline{2-7}
     & $s$ & $p$ & $d$    & $s$ & $p$ & $d$ \\
\hline
Na   & 5550 \newline 6791 
	& 2631 \newline 3356 \newline 4762 \newline 12134 
	& 2820 \newline 3312 \newline 6792 \newline 9961 
	& 6923 \newline 8695 
	& 3269 \newline 4305 \newline 6006 \newline 15386 
	& 3520 \newline 4231 \newline 8567 \newline 12631  \\ 
\hline
Rb   & 5755 \newline 6363 
	& 2735 \newline 3132 \newline 4714 \newline 11836 
	& 2927 \newline 3120 \newline 6743 \newline 9740 
	& 6989 \newline 8358  
	& 3317 \newline 4124 \newline 5934 \newline 15059 
	& 3557 \newline 4095 \newline 8490 \newline 12402  \\ 
\hline\hline
\end{tabular}
\label{table:magic_wlen}
\end{figure}
\end{center}

All these $\lambda^*$ are in the CO$_2$ and the frequency-doubled CO$_2$ laser bands. 
This provides an advantage of individual lattice-site addressing~\cite{SchCatHan00}.  
These wavelengths are far 
from any resonances which reduces photon scattering rate. 

In a trap red-detuned from the Rb $5s-6p$ resonance but  blue-detuned from 
the $5s-5p$ resonance, the ground state polarizability is negative and can be matched to the free-electron polarizability. This allows for a $\lambda^*\approx 432 \, \mathrm{nm}$~\cite{SafWilCla03,SafWalMol10}. However, this magic wavelength being very close 
to the $5s-6p$ resonance  can  lead to enhanced photon scattering and heating.
 Even then, as the``landscape averaging'' reduces the free-electron polarizability employed 
 in~\cite{SafWilCla03}, the feasibility of working at that lattice wavelength needs to be revised. 
 Our calculations show (see Fig.~\ref{fig:Rb_magic}) that for $n=100$ at $\lambda^*\approx 432 \, \mathrm{nm}$, $\alpha^{\rm lsc}_r(\omega^*)/\alpha_e(\omega^*) \approx 3 \times 10^{-3}$, a substantial suppression factor. This reduces the trapping depth and to make up for the suppression the laser intensity would need to be increased by a factor of 360.

In quantum gate protocols such as the CNOT gate, the conditional logic requires driving a 
$\pi$-pulse transition between one of the qubit (ground) states and a Ry state. Although 
both the qubit and the Ry states see the same trapping potentials in magic 
lattices, the differential energy shift between these states does not vanish 
because of the uniform offset term (Eq.~\eqref{eq:pot_offset}). Drifts in the 
lattice laser intensity introduce an error $\Delta\omega$ in the Rabi frequency $\Omega_0$ 
of the GS-Ry transition: $\Omega=\sqrt{\Omega_0^2 + \Delta\omega^2}$. This error in the actual 
Rabi frequency $\Omega$ leads to the fractional error in GS-Ry rotation angle:
$\frac{\Delta\phi}{\pi} \sim
 (\Delta\omega/\Omega_0)^{2}/2$. For Rb, we estimate this error to be 
$\tfrac{\Delta\phi}{\pi} = 25 
	\left(\tfrac{\delta I}{I}\right)^2 
	\left(\tfrac{1\;\text{MHz}}{\Omega_0} \right)^2
	\left(\tfrac{U}{1\;\text{mK}} \right)^2
	\left(\tfrac{\lambda}{1000\;\text{nm}} \right)^4
	\left(\tfrac{350\;\text{a.u.}}{\alpha_{gs}} \right)^2$,
where $\delta I/I$ is the fractional intensity fluctuation and $U$ is the trap depth. For example, 
when $\delta I/I=10^{-4}$ for a 0.16 mK deep trap, $\Omega_0/2\pi=1$ MHz, and  1000 nm lasers, 
$\Delta\phi/\pi = 6.5\times10^{-9}$ . For CO$_2$ wavelengths the errors are below $10^{-4}$, 
which is considered to be tolerable~\cite{Kni05}. 

Finally, although we focused on the magic trapping on the Ry-GS transition, simultaneous magic trapping on the 
qubit transition can be also carried out. 
For example, techniques employing additional compensating CW traveling laser wave~\cite{RadDudZha10} are 
fully compatible with our proposal. Indeed, since the intensity profile of a traveling wave is uniform in 
space, it does not affect the spatially-varying part of  optical potentials.

We have demonstrated that although nominally the Ry state AC polarizability is essentially that of a free 
electron and always negative, laser intensity landscape can profoundly affect the effective 
``landscape-averaged'' polarizabilty and can lead to positive values of  polarizability.  
``Landscape-averaging'' depends on the relative size of Ry orbit and the lattice constant in a 
non-monotonic way. A Ry atom can be attracted to intensity maxima.
This opens up the  possibility of magic trapping of Ry atoms in infrared lattices. 
The separation between adjacent atoms at these IR wavelengths is comparable to Ry blockade 
radius of a few microns, which provides an additional convenience for Ry gate experiments utilizing dipole 
blockade mechanism.

We would like to thank M. Lukin, M. Saffman and E. Tiesinga for discussions. 
This work was supported by the National Science Foundation under Grant No. PHY-1212482 
and PHY05-25915. 


\begin{thebibliography}{10}
\providecommand{\url}[1]{\texttt{#1}}
\providecommand{\urlprefix}{URL }
\providecommand{\eprint}[2][]{\url{#2}}

\bibitem{YeKimKat08}
J.~Ye, H.~J. Kimble, and H.~Katori.
\newblock Science \textbf{320}, 1734 (2008)

\bibitem{TakHonHig05}
M.~Takamoto, F.-L. Hong, R.~Higashi, and H.~Katori.
\newblock Nature \textbf{435}, 321 (2005)

\bibitem{LudZelCam08etal}
A.~D. Ludlow, T.~Zelevinsky, G.~K. Campbell, S.~Blatt, M.~M. Boyd, M.~H.~G.
  de~Miranda, M.~J. Martin, J.~W. Thomsen, S.~M. Foreman, J.~Ye, T.~M. Fortier,
  J.~E. Stalnaker, S.~A. Diddams, Y.~{Le Coq}, Z.~W. Barber, N.~Poli, N.~D.
  Lemke, K.~M. Beck, C.~W. Oates, and G.~K. {Campbell \em et al.}
\newblock Science \textbf{319}, 1805 (2008)

\bibitem{DerKat11}
A.~Derevianko and H.~Katori.
\newblock Rev. Mod. Phys. \textbf{83}, 331 (2011)

\bibitem{RadDudZha10}
A.~G. Radnaev, Y.~O. Dudin, R.~Zhao, H.~H. Jen, S.~D. Jenkins, A.~Kuzmich, and
  T.~A.~B. Kennedy.
\newblock Nat. Phys. \textbf{6}, 894 (2010)

\bibitem{NeyYanMos12}
B.~Neyenhuis, B.~Yan, S.~a. Moses, J.~P. Covey, A.~Chotia, A.~Petrov,
  S.~Kotochigova, J.~Ye, and D.~S. Jin.
\newblock Phys. \ Rev. \ Lett. \textbf{109}, 230403 (2012)

\bibitem{JakCirZol00}
D.~Jaksch, J.~I. Cirac, P.~Zoller, S.~L. Rolston, R.~C\^{o}t\'{e}, and M.~D.
  Lukin.
\newblock Phys. Rev. Lett. \textbf{85}, 2208 (2000)

\bibitem{UrbJohHen09}
E.~Urban, T.~A. Johnson, T.~Henage, L.~Isenhower, D.~D. Yavuz, T.~G. Walker,
  and M.~Saffman.
\newblock Nature \textbf{5}, 110 (2009)

\bibitem{GaeMirWil09}
A.~Ga\"{e}tan, Y.~Miroshnychenko, T.~Wilk, A.~Chotia, M.~Viteau, D.~Comparat,
  P.~Pillet, A.~Browaeys, and P.~Grangier.
\newblock Nat. Phys. \textbf{5}, 115 (2009)

\bibitem{SafWalMol10}
M.~Saffman, T.~Walker, and K.~M\o~lmer.
\newblock Rev. Mod. Phys. \textbf{82}, 2313 (2010)

\bibitem{MorDer12}
M.~J. Morrison and A.~Derevianko.
\newblock \pra \textbf{85}, 33414 (2012)

\bibitem{Der10DoublyMagic}
A.~Derevianko.
\newblock \prl \textbf{105}, 33002 (2010)

\bibitem{Der10Bmagic}
A.~Derevianko.
\newblock \pra \textbf{81}, 051606(R) (2010)

\bibitem{FlaDzuDer08}
V.~V. Flambaum, V.~A. Dzuba, and A.~Derevianko.
\newblock Phys. Rev. Lett. \textbf{101}, 220801 (2008)

\bibitem{ChiNelOlm10}
R.~Chicireanu, K.~Nelson, S.~Olmschenk, N.~Lundblad, A.~Derevianko, and
  J.~Porto.
\newblock \prl \textbf{106}, 063002 (2011)

\bibitem{SafWilCla03}
M.~S. Safronova, C.~J. Williams, and C.~W. Clark.
\newblock Phys.\ Rev.\ A \textbf{67}, 040303(R) (2003)

\bibitem{SafWal05}
M.~Saffman and T.~G. Walker.
\newblock Phys. Rev. A \textbf{72}, 42302 (2005)

\bibitem{ZhaRobSaf11}
S.~Zhang, F.~Robicheaux, and M.~Saffman.
\newblock Phys. \ Rev. \ A \textbf{84}, 043408 (2011)

\bibitem{IvaIsaSaf13}
V.~V. Ivanov, J.~A. Isaacs, M. Saffman, S.~A. Kemme, A.~R. Ellis, G.~R. Brandy, 
J.~R Wendt, G.~W. Biedermann, and S. Samora.
\newblock  	arXiv:1305.5309 [physics.atom-ph] (2013)

\bibitem{YouKnuAnd10}
K.~C. Younge, B. Knuffman, S.~E. Anderson, and G. Raithel.
\newblock Phys. \ Rev. \ Lett. \textbf{104}, 173001 (2010)

\bibitem{DutGueFel00}
S.~K. Dutta, J.~R. Guest, D.~Feldbaum, A.~Walz-Flannigan, and G.~Raithel.
\newblock Phys. \ Rev. \ Lett. \textbf{85}, 5551 (2000)

\bibitem{YouAndGeo10}
K.~C. Younge, S.~E. Anderson, and G.~Raithel.
\newblock New J. Phys. \textbf{12}, 023031 (2010)

\bibitem{OvsDerGib11}
V.~D. Ovsiannikov, A.~Derevianko, and K.~Gibble.
\newblock \prl \textbf{107}, 93003 (2011)

\bibitem{TopDer13}
T. Topcu and A. Derevianko. 
\newblock arXiv:1308.0573 [physics.atom-ph] (2013)

\bibitem{Bloch2012}
I. Bloch, J. Dalibard, and S. Nascimbene, 
\newblock Nat. Phys. \textbf{8}, 267 (2012)

\bibitem{Gal05book}
{T. F. Gallagher}.
\newblock \emph{{Rydberg Atoms}}.
\newblock Cambridge University Press,UK, Cambridge (2005)

\bibitem{LukFleCot01} M. D. Lukin, M. Fleischhauer, R. C\`ote, 
L. M. Duan, D. Jaksch, J. I. Cirac, and P. Zoller, 
\prl {\bf 87}, 037901 (2001)

\bibitem{AmtReeGie07} T. Amthor, M. Reetz-Lamour, C. Giese, and M. Weidem\"uller, 
\pra {\bf 76}, 054702 (2007)

\bibitem{DerJohSaf99}
A.~Derevianko, W.~R. Johnson, M.~S. Safronova, and J.~F. Babb.
\newblock Phys.\ Rev.\ Lett. \textbf{82}, 3589 (1999)

\bibitem{SchCatHan00}
T.~W.~H. {R. Scheunemann, F. S. Cataliotti} and M.~Weitz.
\newblock Phys. Rev. A \textbf{62}, 051801(R) (2000)

\bibitem{Kni05}
E.~Knill.
\newblock Nature \textbf{434}, 39 (2005)

\end{thebibliography}

\end{document}